\documentclass[12pt]{article}
\usepackage{amsmath,amssymb,latexsym,color,graphicx,cite}

\def\R{\Bbb{R}}

\begin{document}
\title{Mathematics of Hybrid Imaging\\ A Brief Review}
\author{Peter Kuchment\\Department of Mathematics\\ Texas A\&M University\\ College Station, TX, 77845, USA\\ kuchment@math.tamu.edu}
\maketitle
\begin{flushright}
\emph{Dedicated to the memory of Professor Leon Ehrenpreis,\\ a great mathematician, human being, and friend.}
\end{flushright}
\date{}
\abstract*{The article provides a brief survey of the mathematics of newly being developed so called ``hybrid'' (also called ``multi-physics'' or ``multi-wave'') imaging techniques.}

\section{Introduction}\label{sec:intro}

Leon Ehrenpreis was a mathematician of remarkable strength, famous accomplishments, and extremely wide interests. In the last couple of decades of his life, integral geometry was one of the main areas he was interested in. This had lead in particular to his involvement with problems of tomography \cite{Leon1,Leon2}. E. T. Quinto and the author had the honor of writing at Leon's request an Appendix \cite{KuQui} dedicated to tomography for his last book \cite{Leon_book}. He was also interested in new developments in medical imaging, which in some instances turned out to be directly related to some integral geometric and PDE problems he considered in \cite{Leon_book}. It is thus appropriate to address some of these new techniques in an article dedicated to Leon's memory.

It is natural to wonder, why do we need new methods of medical imaging in the first place, if we already have the whole bunch of well developed ones \cite{biomed,Natt_old}? Indeed, one can mention the widely known ``standard'' X-ray CT (Computed Tomography) scanners, MRI, and ultrasound scanners, which can be found in most hospitals nowadays. There are also less known to the general public, but well developed by now PET (Positron Emission Tomography) and SPECT (Single Photon Emission Computed Tomography) techniques, Optical Tomography (OT), Electrical Impedance Tomography (EIT), Elastography, as well as quite a few others.

It seems that even the existing methods are way too many. Why would one need all of them? One of the main answers is that different imaging techniques ``see'' different things. One can say (in a crude approximation) that  the CT scan can distinguish between the tissues of different density. What if two tissues (e.g., the cancerous and the healthy one) have essentially the same density, but absorb significantly different amounts of light in certain part of the spectrum? One can think of using OT then rather than the X-ray CT. Some modalities can show the metabolism (e.g., the PET and SPECT), while some others cannot. Some new methods can show the level of oxygenation in blood, while those relying upon density would not be able to do so. This list can go on and on, and so the reader can see the point in having a variety of imaging techniques.

There are several other parameters that make a difference when using different types of scanners. Here are the most common ones:
\begin{enumerate}
\item \textbf{Safety} for the patient and practitioner. Indeed, X-ray scans are clearly not too safe, while for instance OT or ultrasound tomography are not harmful. If there is a choice, one surely would shoot for a safer method.
\item \textbf{Cost}, in the times of the high medical expenses, is clearly one of the major issues. Some imaging techniques (X-ray CT-scan, MRI, PET, and some others) require very expensive devices, with the price tags in millions of dollars. Some others, however, e.g. OT and EIT, are orders of magnitude cheaper.
\item \textbf{Contrast} is another important feature. For instance, if the parameter that a method can detect is, say,  electric conductivity of the tissue, then one will be able to distinguish between the tissues that have a significant conductivity contrast and will not see any difference between tissues that happen to have very close electric conductivities. So, high contrast between the features that we would like to distinguish, is crucial. One clearly would prefer the contrast that is on the order of hundreds (or at least dozens) of percents, while one percent contrast, albeit still usable, is much less desirable. For instance, some breast tumors on early stages might have almost no contrast with the healthy tissues with respect to ultrasound propagation, but a huge contrast (several hundreds of percents) in their optical and electric properties.
\item \textbf{Resolution}, which determines what is the smallest size of a feature that a method can ``see,'' is another very important parameter. Indeed, resolution of several centimeters probably is not good for early breast tumor diagnostics, where sub-millimeter resolution is desirable.
\end{enumerate}

These are the reasons, why the quest for new and ``better'' (at least in some of the parameters) imaging methods not only does not subside, but intensifies in the recent decades, involving new physics, engineering, and mathematics ideas. However, all attempts to find a ``magic wand'' method that would ``see everything'' and feature low cost, safety, high contrast and high resolution, are clearly futile. Thus, having a variety of techniques at our disposal is apparently the way to go.

For a mathematician, however, the main motivation for working on these various tomographic modalities is that they bring about a large variety of challenging and beautiful mathematical problems that involve more or less all areas of mathematics.

Let us describe now the structure of this article.
After a general discussion of hybrid methods in Section \ref{sec:hybrid}, we will concentrate in more details on three of the emerging hybrid methods. The largest Section \ref{sec:TAT} contains an overview of the most developed (both experimentally and mathematically) Thermo-/Photo-acoustic Tomography (TAT/PAT). The next, shorter, Section \ref{sec:AET} addresses the much newer technique of Acousto-Electric Tomography\footnote{Besides the name AET, suggested in the original paper \cite{WangAET}, other names are also used: Ultrasound Modulated EIT (UMEIT) \cite{BalUMEIT}, Impediography \cite{Ammari_book}, and some others.}. Although AET was initially suggested (and its principle experimentally proven) by biomedical engineers \cite{WangAET}, here mathematics is getting developed fast in the recent few years, while experimentalist still struggle with reaching good signal-to-noise ratios (SNR) in the measurements. This situation is reversed in the Ultrasound Modulated Optical Tomography (UMOT, also called UOT), which by now is significantly studied experimentally, while the mathematics of this modality is still in its infancy (and even mathematical models are still being agreed upon). This topic is addressed in ever shorter Section \ref{S:UOT}.

Both AET and UMOT rely upon an assumption of a ``perfect focusing'' of ultrasound at a given location, which is a crude approximation to the reality (see, e.g. \cite{Localized}). The synthetic focusing technique, discussed in Section \ref{sec:focusing}, allows one to use more realistic sets of ultrasound waves.

The common feature of AET and UMOT (as well as some other hybrid imaging techniques, such as CDI and MREIT, which are just mentioned in this text) is that the measurements provide the researcher with some interior information, i.e. a function of an interior location $x$. There has been a rather common feeling that such an interior information might stabilize the notoriously unstable modalities such as EIT or OT. This issue is briefly discussed in Section \ref{sec:interior}.

The topics surveyed in this article are highly technical and involve a wide range of interesting mathematics. Due to space limitations, the author was forced to show very few technical details and instead to try to give a hand-waving heuristic description. The literature references (as well as the references therein) will provide the interested reader with more details.

\section{Hybrid imaging methods}\label{sec:hybrid}

As we have indicated in the introduction, each of the available imaging methods has its advantages and deficiencies. For instance, in breast imaging ultrasound provides a high (sub-millimeter) resolution, while suffers from a low contrast. On the other hand, many tumors absorb much more energy of electromagnetic waves (in some specific energy bands) than healthy cells. Thus, using such electromagnetic waves offers very high contrast. Alas, the resolution in this case is very low. One can go on and on with such examples.

Since both the advantages and disadvantages of various modalities come from the underlying physics, can one do anything to
combine the advantages and simultaneously alleviate the problems associated with different type of physical waves/radiation involved? The natural idea is to try to this end to combine different imaging modalities into some kind of ``hybrid'' ones. This is how the \textbf{hybrid} (also called \textbf{multi-physics} or \textbf{multi-wave}) \textbf{techniques} have been appearing in the last decade - decade and a half \cite{Ammari_book,ScherSpring,Tuchin,vodinh,IOut2,Banff,WangPATbook,WangTextbook,BalHybr}.

The ``hybridization'' can occur at different stages of the imaging process. Let us recall the crude scheme of all CT methods: on the first step, some wave(s) are sent through the body and the outgoing (transmitted or reflected) waves are measured; on the second stage, mathematical processing of the measured data is done; finally, the third stage provides a picture (tomogram). Correspondingly, one can combine different techniques at different stages. The simplest, and already industrially implemented (e.g., in PET/CT scanners) idea is to run two types of scans of the same patient and then somehow ``combine'' the images so they hopefully complement each other. Here reconstruction of both individual images requires neither new types of scanners, nor new mathematical reconstruction algorithms. Certainly, some (often non-trivial) processing is needed to correctly overlap the two images (the so called \textbf{image registration}).

Another option is to combine the techniques on the second stage. Namely, after collecting data from two independent types of scans, a reconstruction algorithm is applied that uses both data sets. The additional information can significantly improve the quality (stability, resolution, etc.) of the resulting picture (tomogram). Such combined techniques might not require any new physics or engineering, but do demand new mathematical processing. There are several recent imaging procedures that successfully implement this idea. Some of them (CDI - current density imaging \cite{Nach1,Nach2,Nach3} and MREIT \cite{SeoSIAM,SeoMREIT}) combine MRI and EIT measurements. Having the extra MRI data makes the mathematical problem of EIT reconstruction (the so called Calder\'{o}n problem, or inverse conductivity problem \cite{SylvUhl,BB1,Bor02,Cald,CIN,Kenig,Nach,AstalaPaiv,biomed,Uhlm_asteris,IOut1,UhlmCald}), known for its severe ill-posedness, significantly less ill-posed and thus allows for good quality reconstructions of the internal electrical conductivity maps. Another actively developing method of this kind, called MRE \cite{MclYoon,MRE}, combines MRI with elastography: MRI allows one to observe propagation of elastic waves through the tissue, which then leads to mathematical reconstruction of mechanical properties (e.g., stiffness) that carry important medical diagnostic information. Probably the oldest such combination is of CT and SPECT. The CT scan provides the reconstruction of the attenuation map, which is used then to recover the distribution of a radio-pharmaceutical inside the patient's body \cite{BGH}.

Due to the author's limited expertise and lack of space, these types of hybrid methods will not be discussed in the article. The interested readers are referred to the literature cited above.

We reserve in this text the name \textbf{``hybrid methods''}
only for the techniques that combine different types of waves already on the first, scanning stage. In the examples that we will discuss this will lead to one physical type of irradiation triggering or modulating the other one and thus producing new types of measurements, which hopefully allow one to improve images in comparison with the two techniques done separately\footnote{The reader should be aware that this classification of hybrid modalities into three classes, although being reasonable, is not commonly accepted and is used in this text only for the author's convenience.}.

We now move to considerations of some of the hybrid techniques in more detail. We start in the next section with the probably best developed (both experimentally and mathematically) among the hybrid techniques Thermo-/Photo-acoustic tomography (TAT/PAT) and then move to the less studied ultrasound modulated electrical and optical tomography.

\section{Thermo-/Photo-acoustic tomography (TAT/PAT)}\label{sec:TAT}

As we have already mentioned, in many medical diagnostics situations, ultrasound displays low contrast (and thus sees the tissues as almost homogeneous), while providing fine resolution. Optical or radio-frequency EM illumination, on the other hand, gives an enormous contrast between the cancerous and healthy cells, while both are known to suffer from low resolution. How can one combine their strengths? The answer is in the photo-acoustic effect, which was discovered by Alexander Graham Bell \cite{Bell}, but had to wait for another century for its applications to follow \cite{Bowen}.

\begin{figure}[ht!]
    \includegraphics[scale=0.7]{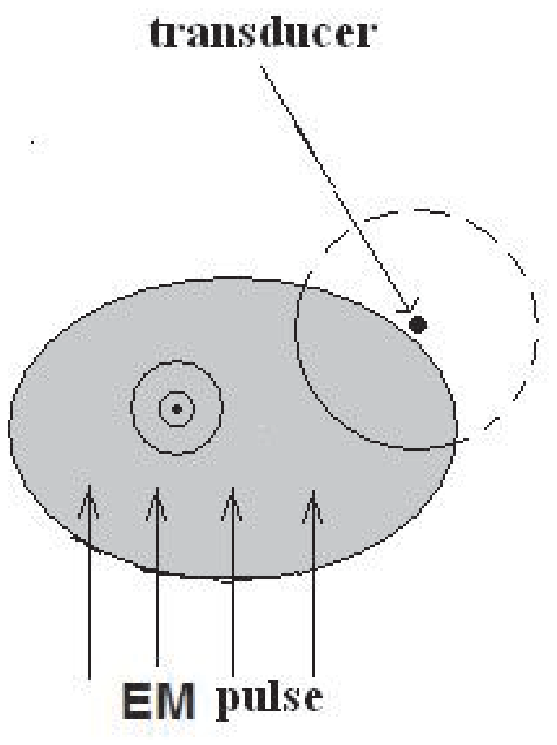} \includegraphics[scale=0.7]{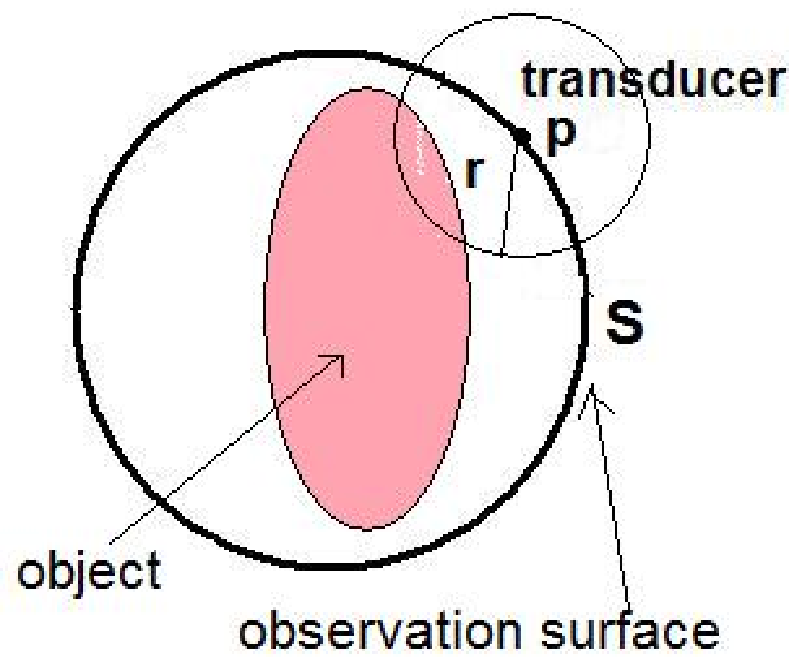}
\caption{The TAT procedure.}\label{F:TAT}
\end{figure}

Imagine that a biological object is irradiated by a wide, homogeneous, but extremely short electro-magnetic pulse in radio-frequency range (Fig. \ref{F:TAT}).
Some part of the electromagnetic (EM) energy will be absorbed throughout the tissues. Let us denote by $f(x)$ the density of energy absorbed at a location $x$. It is known that the values of this absorption function will be several times higher at the cancerous locations than in the surrounding healthy tissues \cite{Diebold,AE2,AE3,Oraev94,Oraev96,Tam,WangPATbook,WangTextbook,Anast_hand,MXW_review,Tuchin,vodinh}. Thus, if this function were known, it would provide an extremely valuable diagnostic information. However, the radio frequency waves are too long to lead to any reasonable resolution. This is where the photo-acoustic effect kicks in: the EM energy absorption leads to heating, and higher levels of absorption lead to more heating. In turn, the resulting thermoelastic expansion creates a pressure wave $p(x,t)$ (acoustic wave), whose initial value is essentially (proportional to) the function of interest $f(x)$. Placing an ultrasonic transducer (a microphone) at a location $y$ at the boundary of the object, one can measure the value of $p(y,t)$ at this point for any time $t\geq 0$. If we now surround the object by transducers located on an \textbf{observation surface} $S$ (this can be done using optical interferometers, which do not obstruct the acoustic wave propagation), we can collect the values of $p(y,t)$ for all $(y,t)\in S\times\R^+$ (Fig. \ref{F:TAT}). Now the problem reduces to finding the initial values $p(x,0)$ inside the volume surrounded by the surface $S$. This is the idea of Thermoacoustic Tomography (TAT), which found its implementation in the middle of 1990s \cite{Kruger,Oraev94,Oraev96}, in particular in the work of R. Kruger \cite{Kruger}, who started a company OptoSonics manufacturing TAT devices. The Photo-acoustic version (PAT) differs by the choice of heating radiation - a laser beam instead of radio-waves. A large part of the corresponding mathematics is parallel in TAT and PAT cases, and thus we will only mention TAT here.

The mathematics of TAT/PAT reconstruction happens to be fascinating and occupied attention of a large group of mathematicians throughout the 1st decade of this century. The reader can consult with the recent surveys, collections, and books \cite{AKK,AmmariPAT,Banff,FR,FR2,KuKuEJM,KuKuElectr,Pal_book,PatSch,ScherSpring,KuScher,StUhl,Tuchin,vodinh,IOut2} for details and further references.

We will now describe the mathematical model of TAT. Let us denote by $c(x)$ the sound speed in the interior of the body. Then the pressure $p$ satisfies the following wave equation:
\begin{equation}\label{E:wave}
\begin{cases}
\frac{\partial^2 p}{\partial t^2}=c^2(x)\Delta p,\quad \mbox{ in }\R^3\times\R_+\\
p(0,x)=f(x),\\ \frac{\partial p}{\partial t}(0,x)=0 \\
\end{cases}
\end{equation}
The data $g$ measured by a TAT machine provides the values of the pressure $p$ on the observation surface $S$:
\begin{equation}\label{E:wave_data}
f(x)\mapsto g(y,t):=p\mid_{S\times\R_+}.
\end{equation}
The goal of TAT thus is inverting this forward operator.

\subsection{TAT/PAT and (restricted) spherical mean operators}

In the case of an acoustically homogeneous medium (i.e., when $c(x)=$\emph{const}), one can reduce the TAT problem to an equivalent integral geometry question. Namely, using the standard Kirchhoff-Poisson formulas \cite{CH,John} for the solution of (\ref{E:wave}), the TAT inversion can be reduced to the equivalent (see \cite{AQ}) problem of recovery the function $f(x)$ from its spherical averages over spheres centered at the transducers' locations, i.e., on the observation surface $S$:
\begin{equation}\label{E:averages}
f\mapsto g(y,t):=(R_Sf)(y,t)=\int_{|x-y|=t} f(x)d\sigma(x), y\in S, t\geq 0.
\end{equation}
Here $R_S$ denotes the operator of taking spherical averages over all spheres centered on $S$.

One can notice that such transforms were considered in more general situation and without any relation to tomography by Leon Ehrenpreis in his book \cite{Leon_book} and by V.~Lin and A.~Pinkus in \cite{LP1,LP2} for the needs of approximation theory and neural networks. The restricted spherical mean transforms also play important role in the Radar and Sonar studies \cite{LQ,Cheney}. One can find a list of other applications in \cite{AQ}.

\subsection{Main mathematical problems in TAT}

As in all tomographic methods, the following questions play the central role:
\begin{description}
\item[\textbf{\underline{Uniqueness of reconstruction:}}] Is the collected data $g$ sufficient for the unique reconstruction of the tomogram (function $f(x)$)? When the author first looked at this problem in terms of the system (\ref{E:wave}), he was confused for a second. Indeed, the measured data $g$ seems to be the boundary value of the solution of the wave equation in a cylinder, and the function $f$ to recover is the initial condition. This clearly is impossible, since $g$ essentially does not carry any information about $f$, besides the standard junction conditions where these functions meet. However, the thing is that (\ref{E:wave}) is not a boundary value problem in a cylinder, but rather a problem in the whole space, whose solution we observe on the surface $S$ only. What is even more important, if the sound speed is non-trapping (e.g., constant) and $S$ is a closed surface (i.e., the boundary of a bounded domain), then the local energy decay theorems \cite{Egorov,Vainb,Vainb2,Ral,Burq} show that the solution decays in the interior of $S$. This turns out to be the main feature that brings about uniqueness and inversion \cite{AK,Hristova,HKN,StUhl}.

    The question of sufficiency of the data collected on a non-closed hyper-surface $S$ is much more complicated and can be considered largely resolved only in $2D$, when the sound speed is constant and the function $f(x)$ is compactly supported (see \cite{AQ}\footnote{Some microlocal arguments in \cite{AQ}, although correct, were incomplete. The missing arguments are provided in \cite{StefUhl}.}, as well as surveys \cite{AKK,KuKuEJM,KuKuElectr} for the results and further discussion). This is a fascinating mathematical problem, which involves a variety of techniques from PDEs, integral, differential, and algebraic geometry, microlocal analysis, commutative algebra, etc.
    A more general problem (with spheres replaced by the level surfaces of a polynomial) is discussed in L. Ehrenpreis' book \cite{Leon_book}.

\item[\textbf{\underline{Reconstruction procedures:}}] Having a uniqueness of reconstruction theorem is far from being sufficient for tomography, since one needs to be able to actually reconstruct the tomogram. So, the natural question, after proving uniqueness, is: how can one actually invert the mapping $f\mapsto g$? We will briefly describe the known algorithms in the next sub-section.

\item[\textbf{\underline{Stability:}}] Having uniqueness of reconstruction theorem, or even a reconstruction algorithm is no guarantee that the reconstruction is practicably feasible (or at least that it can provide sharp pictures). The word ``stability'' in this context means the effect that small errors in the measured data have on the reconstructed image. In other words, stability means well-posedness (in the Hadamard's sense \cite{CH}) of the inverse problem. Unfortunately, essentially all tomographic problems are ill-posed to some degree. Some of them, such as Radon transform inversion, are very mildly unstable, which allows for wonderful CT-scan images. Some others, like EIT and OT, are known to be severely ill-posed, and thus reconstruction of sharp images is practically impossible.

    The stability of the TAT/PAT reconstruction with full data (i.e., for the observation surface surrounding the object completely) is known to be very good, the same as for the Radon transform inversion, which leads to excellent reconstructions. This applies both to the case of an acoustically homogeneous object (when $c(x)=$ \emph{const}), as well as to the case of a variable non-trapping sound speed $c(x)$ \cite{AK,HKN,nguyen_stab,StUhl,Dustin_uniq,Dustin_req}. The proofs are based upon inversion procedures and/or a microlocal analysis of the problem.

\item[\textbf{\underline{Range:}}] As it is common in integral geometry and tomography \cite{Natt_old,Natt_new,Helg_Radon,Helg_new,GelfVil,GGG}, the range of the forward operator $f \mapsto g$ has infinite co-dimension in the natural scales of function spaces. The description of this range is an important part of integral geometric and tomographic studies [\emph{ibid}]. In the case of a spherical observation surface $S$ and constant sound speed, the complete range descriptions are known  \cite{AKQ,AFK,Finch_range,AmbKuc_rang,AL}. Some of these range conditions are known to be necessary for more general observation surfaces and sound speeds, but in these cases complete range description is not known (and might be impossible).

\item[\textbf{\underline{Incomplete data:}}] In the TAT/PAT case, one usually mentions an \textbf{incomplete data situation}, when the observation surface $S$ does not surround the object completely \cite{AKK,KuKuElectr,KuKuEJM,XWAK,XWAK2}. This is a common, albeit somewhat misleading description, since depending on the location of the object, the ``incomplete'' data might be sufficient for unique, and sometimes even stable reconstruction. It is quite common that even a small observation surface $S$ can provide enough data for proving uniqueness of reconstruction. For example, in the case of a constant sound speed, it is known that any set $S$ that is not a part of an algebraic hyper-surface, leads to the injectivity of the spherical mean operator $R_S$, and thus to unique TAT reconstruction (see \cite{AQ} and references therein). Note that such $S$ can be geometrically very small. It is clear that for all practical purposes, in spite of an uniqueness theorem, something should go wrong with the actual reconstruction in this case. And indeed, microlocal arguments similar to those of X-ray CT \cite{Quinto} show that most of the singularities (i.e., wave front set directions) of $f(x)$ will be ``invisible'' (``not audible'')  \cite{AKK,KuKuElectr,KuKuEJM,XWAK,XWAK2,nguyen_stab,StUhl,HKN,Palam_funk,Pal_book}. This implies, in particular, the high instability of the reconstruction (all ``invisible'' interfaces will be blurred, no matter how sophisticated the algorithms are) \cite{XWAK,XWAK2,nguyen_stab,StUhl,HKN}. This is also well known in the more standard X-ray CT and SPECT \cite{Quinto,KLM}. The wonderful feature of the microlocal analysis is that it not only explains, but also allows one to predict these blurring effects [\emph{ibid}].

    On the other hand, if one is in the situation where there is a uniqueness result and all singularities of the object are ``visible'', one indeed can reconstruct the object stably \cite{Kun_open,Anastasio_halftime,PopSuch_half}. Conditions of unique reconstruction and ``visibility'' have been also figured out for the case of variable sound speeds and can be expressed simply in terms of geometric optics rays \cite{AKK,KuKuElectr,KuKuEJM,XWAK,XWAK2,nguyen_stab,StUhl,HKN}.

\end{description}

While the questions above are common for all tomographic techniques, there is one issue that is specific for TAT (a similar, still not completely resolved, problem arises also in SPECT, see \cite{Kuch_AMS05} and references therein, as well as the recent papers \cite{BalJollSPECT,Bukhgeim}). This is the
\begin{description}
\item{\underline{\textbf{Speed recovery problem:}}}
The problem (\ref{E:wave}) involves the unknown function $f(x)$, the tomogram, as well as the sound speed $c(x)$, which in all reconstruction methods is assumed to be known. It has been observed \cite{HKN,JinWang} that errors in the values $c(x)$ introduce significant artifacts into reconstruction. In other words, one needs to know the speed $c(x)$ well, which is normally not the case. One of the options suggested to alleviate this difficulty, is to run an ultrasound transmission scan beforehand, which would provide the speed map \cite{JinWang}. However, there is numerical evidence \cite{Anastasio_speed,Zhang} that it might be possible to determine both the speed $c(x)$ and then the tomogram $f(x)$ from the TAT data. Proving uniqueness of reconstruction of $c(x)$ happens to be a difficult problem. Some very limited partial results have been obtained recently (and not published, except \cite{Hickmann,HKN}) by various mathematicians: M.~Agranovsky, D.~Finch, K.~Hickmann, Y.~Hristova, P.~Kuchment, L.~Nguyen, P.~Stefanov, and G.~Uhlmann. However, the problem (which, according to D. Finch's observation, is closely related to also not completely resolved well known transmission eigenvalue problem \cite{TransEig,Kirsch,ColtonKress,ColtonTransEig,PaivSyl,Sylv,Cakoni}) is essentially open. Somewhat more understanding has been achieved concerning possible instability of the speed reconstruction (L.~Nguyen, P.~Stefanov, G.~Uhlmann, unpublished).
\end{description}

\subsection{Reconstruction methods in TAT/PAT}
What techniques are available for actual TAT reconstructions? There are several groups of approaches that have been used successfully:

\begin{description}

\item[\underline{\textbf{Closed form inversion formulas:}}]  Closed-form inversion formulas in tomography bring
about better theoretical understanding and lead to efficient reconstruction algorithms. The best known example is the so-called
filtered backprojection (FBP) algorithm in X-ray tomography, which is derived
from one of the popular inversion formulas for the classical Radon transform (see, for
example \cite{Natt_old,Natt_new,Helg_new,Helg_Radon,Herman}).

The first such formulas for TAT (the very existence of which had not been clear) were obtained in odd dimensions in \cite{FPR}, for the observation surface $S$ being
a sphere and the sound speed being constant. Let the function $f(x)$ be supported within a ball of radius $R$ and the detectors be located on the boundary $S=\partial B$ of
this ball. Some of the formulas obtained in \cite{FPR} are:
\begin{align}
f(x)  &  =-\frac{1}{8\pi^{2}R}\Delta_{x}\int\limits_{\partial B}
\frac{g(y,|y-x|)}{|y-x|}dA(y),\label{FPR3da}\\
f(x)  &  =-\frac{1}{8\pi^{2}R}\int\limits_{\partial B}\left(  \frac{1}{r}
\frac{\partial^{2}}{\partial r^{2}}g(y,r)\right)  \left.
{\phantom{\rule{1pt}{8mm}}}\right\vert _{r=|y-x|}dA(y),\label{FPR3d}\\
f(x)  &  =-\frac{1}{8\pi^{2}R}\int\limits_{\partial B}\left(  \frac{1}{r}
\frac{\partial}{\partial r}\left(  r\frac{\partial}{\partial r}\frac
{g(y,r)}{r}\right)  \right)  \left.  \phantom{\rule{1pt}{8mm} }\right\vert
_{r=|y-x|}dA(y), \label{FPR3db}
\end{align}
where $dA(y)$ is the surface measure on $\partial B$ and $g$ represents the
values of the spherical integrals (\ref{E:averages}).

Here
differentiation with respect to $r$ in (\ref{FPR3d}) and (\ref{FPR3db}) and
the Laplace operator in (\ref{FPR3da}) represent the filtration step, while the
(weighted) integrals correspond to the backprojection: integration over
the set of spheres passing through the point $x$ and centered on $S$.

A different family of inversion
formulas valid in any dimension was found in
\cite{Ku2007}. Still another set of closed-form inversion formulas applicable in even dimensions was found in \cite{Finch_even}.
Finally, a unified family of inversion formulas was derived in \cite{Linh}.

Having closed form inversion formulas has the advantage that they usually lead to fast and precise inversion algorithms. However, there are several disadvantages of these formulas in TAT. First of all, the FBP formulas are now available only for the observation surface being a sphere (see discussion above), with the only exception of \cite{Kun_cube}, where such formulas are derived for a cube and some other crystallographic domains. Secondly, it is known (e.g., \cite{KuKuEJM}) that if some part of the source function $f(x)$ is supported outside the observation surface $S$, then its reconstruction inside $S$ using FBP formulas might be incorrect. Finally, there are no FBP formulas known for the case of a variable sound speed.

\item[\underline{\textbf{Eigenfunction expansions:}}] -
    This approach, which theoretically works for arbitrary closed surfaces, was proposed in \cite{Kun_series} (and extended in \cite{AK} to the case of variable sound speeds). It is based on expansion into eigenfunctions of the Laplacian operator in the interior $B$ of $S$ with zero Dirichlet conditions on $S$. It is thus nicely implementable whenever the spectrum and eigenfunctions of the Dirichlet Laplacian are known explicitly, e.g., for spheres, half-spheres, cylinders, cubes and parallelepipeds, as well as the surfaces of some crystallographic domains.

    The function $f(x)$ can be reconstructed inside $B$ from the data $g$ in (\ref{E:wave}), as the following $L^{2}(B)$-convergent series:
        \begin{equation} f(x)=\sum\limits_{k}f_{k}\psi_{k}(x), \label{E:coef_variable}
    \end{equation}
    where $\psi_k(x)$ are properly normalized eigenfunctions of the operator $(-c^2(x)\Delta)$ in $B$ with zero Dirichlet conditions and $\lambda_k^2$ are the corresponding eigenvalues. The Fourier coefficients $f_{k}$ can be recovered from the data $g$ in (\ref{E:averages}) using one of the following formulas: \begin{equation}
    \begin{array} [c]{c} f_{k}=\lambda_{k}^{-2}g_{k}(0)-\lambda_{k}^{-3}\int\limits_{0}^{\infty} \sin{(\lambda_{k}t)}g_{k}^{\prime\prime}(t)dt,\\ f_{k}=\lambda_{k}^{-2}g_{k}(0)+\lambda_{k}^{-2}\int\limits_{0}^{\infty} \cos{(\lambda_{k}t)}g_{k}^{\prime}(t)dt,\mbox{ or }\\ f_{k}=-\lambda_{k}^{-1}\int\limits_{0}^{\infty}\sin{(\lambda_{k}t)} g_{k}(t)dt=-\lambda_{k}^{-1}\int\limits_{0}^{\infty}\int\limits_{S} \sin{(\lambda_{k}t)}g(x,t)\overline{\frac{\partial\psi_{k}}{\partial n} (x)}dxdt,
    \end{array}
    \label{E:coef_variable2}
    \end{equation}
    where \[
g_{k}(t)=\int\limits_{S}g(x,t)\overline{\frac{\partial\psi_{k}}{\partial
n}(x)}dx.
\]

    This method becomes computationally efficient when the eigenvalues and eigenfunctions are known explicitly and a fast summation formula for the series (\ref{E:coef_variable}) is available, for instance when the acquisition surface $S$ is a surface of a cube, and thus the eigenfunctions are products of sine functions. The resulting $3D$ reconstruction algorithm is extremely fast and precise (see~\cite{Kun_series}).

This method applies in any dimension and is stable. It also does not have the deficiencies of the FBP formulas that we have mentioned above. Namely, presence of a part of the function $f(x)$ outside $f(x)$ does not hurt the reconstruction inside. The method, at least theoretically, works for arbitrary closed observation surface $s$ and variable speed $c(x)$. However, its practicality in these circumstances is still questionable.

\item[\underline{\textbf{Time reversal:}}] -
We now describe an inversion technique that has the same advantages as the eigenfunction expansion method above, and in addition is easy to implement for any shape of the observation surface and acoustically inhomogeneous media. One can come up easily with this method, if one notices the underlying assumption of TAT, which is often hidden, and which we have discussed explicitly before: local energy decay of the solution of (\ref{E:wave}). Then one can naturally come up with the idea of running the wave equation in  (\ref{E:wave}) back in time, starting at the infinite time with zero initial condition (which reflects the local energy decay) and using the measured data $g$ as the boundary value. Eventually, at time $t=0$, one arrives to the tomogram $f(x)$. Certainly, the solution never vanishes exactly at any finite time (unless the sound speed is constant and the dimension is odd, where the Huygens' principle kicks in). Thus, one has to start from some sufficiently large time $t=T$ with zero conditions and go back in time to arrive to an approximation of $f(x)$, which one expects to get better when $T\to\infty$. This approach has been implemented by various researchers, its feasibility was shown, and error estimates were provided (see, e.g., \cite{Wang_reversal,Grun,HKN,Hristova}). This works \cite{Grun,HKN} even in $2D$ (where decay is the slowest) and in inhomogeneous media. However, when trapping occurs, some parts become ``invisible'' and blur away \cite{HKN}.

A more sophisticated (than just zero) cut-off at time $t=T$ is used in the version of time reversal suggested in \cite{StUhl,StUhl2,Qian}. This leads to an equation with a contraction operator, which allows the use of the Neumann series (fixed point iterations) to obtain high quality images.

It is the author's belief that the time
reversal is the most versatile and easy to implement method of TAT reconstructions.

\item[\underline{\textbf{Algebraic reconstruction techniques:}}] ART is the well used workhorse in approaching inverse problems (especially, when analysis of the problem is too complicated). To put it simply, in ART one discretizes the problem and uses one's favorite method (usually an iterative one) for solving the resulting linear algebraic system. Such techniques have been used in CT for quite a while
\cite{Natt_old}.
ART algorithms frequently produce very good images. However, they are notoriously slow. In TAT, they have been used successfully for reconstructions with partial data
\cite{Paltaufiter,Anastasio_halftime} and sound speed recovery \cite{Zhang,Anastasio_speed}.

\item[\underline{\textbf{Parametrix approaches:}}]
    Some of the earlier non-iterative reconstruction techniques of approximate nature \cite{Kruger,PopSush,PopSush2} were based (explicitly, or implicitly) upon microlocal analysis. For example, in \cite{Kruger}, by  approximating the integration spheres
by their tangent planes at the point of reconstruction and applying one of
the inversion formulas for the classical Radon transform, one
reconstructs a decent approximation to the image.

Such techniques are related to the general scheme proposed in \cite{Be}
for the inversion of the ``generalized'' Radon transform that
integrates over curved manifolds.
One constructs a \textbf{parametrix} (usually an integral Fourier operator (FIO)) for the forward operator $F:f\to g$, i.e. such operator $P$ that the operators $PF-I$ and $FP-I$, while not equal to zero, as in the case of a true inversion, are ``smoothing.'' Thus, applying a parametrix $P$ to the data $g$, one recovers the image $f$ up to addition of a smooth function. This also often reduces the problem to a Fredholm
integral equation of the second kind, which is well amenable to numerical
solution. In other words, the parametrix method provides an efficient pre-conditioner for an iterative solver; the convergence of such iterations can be much faster than that of algebraic iterative methods. On the other hand, parametrix reconstructions can be often accepted as approximate images.

\end{description}

\subsection{Examples of TAT/PAT reconstructions}
\begin{figure}[ht!]
       \includegraphics[scale=0.6]{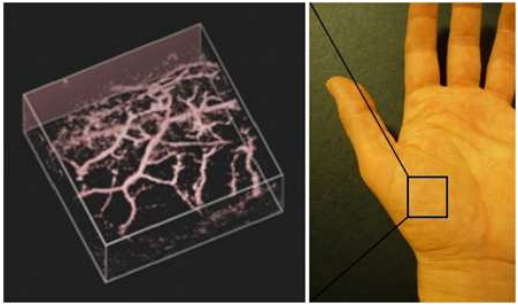}
   \caption{A PAT reconstruction of the blood vessel system of a human hand. The picture is courtesy of Wikipedia.}\label{F:reconstr}
\end{figure}

Now, the mathematics looks nice, but does the TAT/PAT procedure really work? The reader can find below  (Fig. \ref{F:reconstr}) the wonderful PAT image (curtesy of Wikipedia) of a blood vessel structure inside a human hand.

The next Fig. \ref{F:partial} (from \cite{XWAK}) shows several TAT reconstructions of a muscle+fat phantom shown on top left.

\begin{figure}[ht!]
    \includegraphics[scale=0.6]{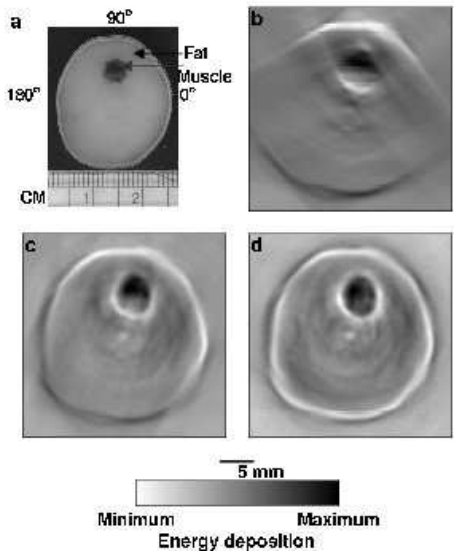}
\caption{a. The phantom. b. Reconstruction from partial data with ``invisible'' details blurred. c. Reconstruction from partial data with all features visible. d. Reconstruction from full data \cite{XWAK}.}\label{F:partial}
\end{figure}

\subsection{Quantitative PAT}

The usual TAT/PAT procedure essentially recovers the initial pressure $f(x)$, which is proportional to the energy deposition function $H(x)$.  How does $H(x)$ relate to the actual electric or optical parameters of the medium?
It is less of a problem for TAT, where radio frequencies are used and one can hope to obtain more or less homogeneous irradiation of all tissues. The situation is different in PAT.
In the diffusion regime, one has
$$
H(x)=\Gamma (x) \sigma(x) u(x),
$$
where $\Gamma$ is the so called Gr\"{u}neisen coefficient, $\sigma(x)$ is the EM energy absorption coefficient, and $u(x)$ is the radiation intensity. The following equation is satisfied by $u(x)$:
$$
-\nabla\cdot D(x)\nabla u(x)+\sigma(x) u(x)=0,
$$
where $D$ is the diffusion coefficient. The question arises whether, after doing the TAT/PAT reconstruction and recovering $H(x)$, one can go further and recover the actual optical parameters $(D, \sigma, \Gamma)$ from $H(x)$? This is the goal of the so called Quantitative PAT (QPAT), which has started developing only recently \cite{BalJoll,CoxQPAT,BalRen,BalRenQPAT,Zhang,RenBal,Cox3:2009,Cox4:2009,BalUhl} and is in a very active stage now. In other words, QPAT takes of where PAT lands.

\section{Acousto-Electric Tomography}\label{sec:AET}
Electrical impedance tomography (EIT) strives to recover the interior distribution of electric conductivity by measurements on the boundary \cite{BB1,Bor02,Cald,CIN,Cipra,Kenig,Nach,AstalaPaiv,biomed,Uhlm_asteris,IOut1,UhlmCald,SylvUhl}. Namely, one creates various boundary voltages and measures the resulting boundary currents (or vice versa). From these measurements one tries to recover the internal conductivity. The mathematical incarnation of EIT is the \textbf{inverse conductivity problem}, which was apparently suggested first by E. Calder\'{o}n and has by now a glorious 30 years history.  Efforts of many leading mathematicians were directed towards proving that the measured data is sufficient for the unique recovery of internal conductivity. This happens to be much more mathematically difficult topic than those arising in traditional tomography or in TAT. To large extend, this is due to the significant non-linearity of the corresponding mapping and instability of its inversion. This is still an active area, for instance since the optimal result in $3D$ is still not known. However, the general understanding is that one does have sufficient information for the reconstruction of the conductivity. We provide just a sample of references to this bursting with life topic \cite{Bor02,Cald,CIN,Kenig,AstalaPaiv,Uhlm_asteris,IOut1,UhlmCald,SylvUhl}, where the reader can find plenty of information and further references.

\begin{figure}[ht!]
    \includegraphics[scale=0.6]{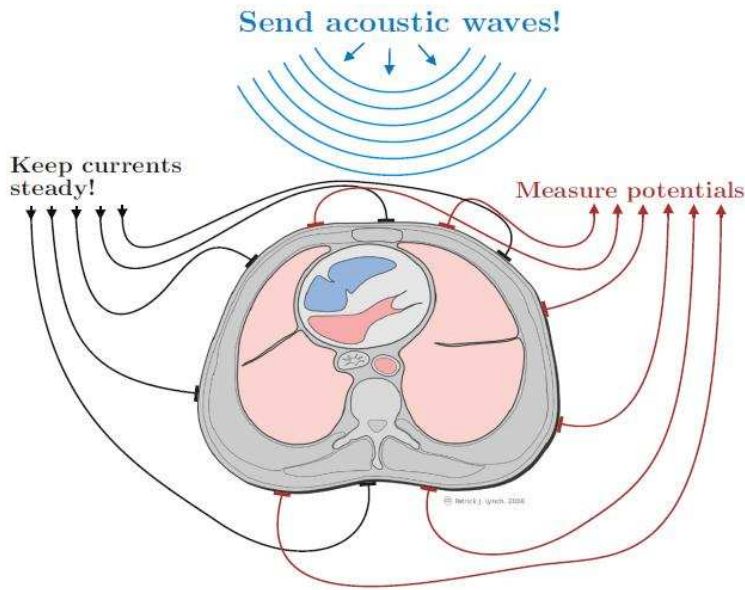}
\caption{The AET procedure: electrical boundary measurements are done concurrently with scanning the object with ultrasound. The picture is courtesy of L. Kunyansky.}\label{F:AET}
\end{figure}

In this text we will be interested in the issues
related to the actual reconstruction in EIT, which has also attracted enormous attention of scientists, including mathematicians. Due to the already mentioned instability\footnote{Here instability = ill-posedness = exponential decay of singular values of the direct mapping.}, the pictures come out very much blurred and of low resolution. This is a rigorously proven fact of life, not the deficiency of the mathematics used. It is the author's opinion that by now experts have achieved as good EIT reconstructions as humanly possible ... unless one changes the physical set-up of the measurements and/or uses some \emph{a priori} information. Changing the experimental set-up is what is suggested in Acousto-Electric Tomography (AET).

The main obstacle of EIT (and similarly of OT (optical tomography)) is that the signals measured at the boundary loose, exponentially fast with the depth, the information on where they came from. This is a simple minded explanation of why reconstructions are blurred. So, if one could somehow get some type of ``interior'' information about where the signals came from, one could hope for stabler image reconstructions. AET is one of the implementations of this idea.
It has been known for some time that ultrasound irradiation of soft tissues modifies the tissues' electric and optical properties (electro-acoustic effect \cite{AE2,AE3}). It was thus an easy step to decide to send an ultrasound beam that focuses on some internal location $x$ and thus modifies (by a multiplicative factor close to $1$) the electric conductivity $\sigma(x)$ at this location. This would lead to a perturbation of the boundary EIT measurements, and what is crucial, the practitioner will know where the perturbation came from - from the point $x$. Then one could scan the focused beam throughout the whole object and get hopefully sufficient information for a stable reconstruction. This idea of AET was suggested and tried by a direct measurement in \cite{WangAET}. It was shown there that a detectable (albeit rather small) signal does exist. However, no reconstruction was done at that time. In a few years, the topic started developing fast \cite{WangAET,Ammari_book,Ammari_AET,bon,Cap,ScherAET,KuKuAET,KuKuSynt}, sometimes with the researchers being unaware of the original work \cite{WangAET}. Let us describe briefly the current state of affairs in AET (although by the time of publication, the situation will definitely change).

The following observation was made experimentally and justified theoretically \cite{AE2,AE3,WangAET}: the acousto-electric effect, although detectable, is so small, that one can safely linearize the problem.

Another smallness assumption was used in most works: the ability of sharp focusing at a given location (i.e., creating a delta-type ultrasound pulse). Such perfect focusing is clearly impossible (see the discussion in the book \cite{Localized} devoted to this issue). Still, let us assume for the time being that sufficiently good focusing is possible (and return to this discussion in Section \ref{sec:focusing}). This allows one to use the well studied ``small volume inclusion'' asymptotics, as in \cite{Ammari_book,Ammari_AET,Cap,Vogelius}, where such asymptotics play the crucial role. On the other hand, in \cite{KuKuAET,KuKuSynt}, only smallness of the acousto-electric effect is needed, and no perfect focusing is required (see Section \ref{sec:focusing}).

In all these works the authors studied what kind of interior quantities can be stably recovered from the measurements, if perfect focusing (in particular, small volume asymptotics) is possible. For instance, if $u_1(x), u_2(x)$ are the (unknown) potentials created by some boundary current set-up, then one can recover the values
$$
\sigma(x) \nabla u_1(x)\cdot \nabla u_2(x)
$$
for any interior point $x$, where $u\cdot v$ denotes the inner product of two vectors. Some other local functionals of the form $F(\sigma(x), u_i(x), \nabla u_i(x))$ could be recovered (this is also the case in the previously mentioned MREIT and CDI \cite{Nach1,Nach2,Nach3,SeoSIAM,SeoMREIT}, which we do not discuss in this paper).

It was shown then that such values, if recovered from measurements, lead to locally unique and stable reconstruction of the conductivity $\sigma$ \cite{KuKuAET,KuKuSynt,bon,Cap,Dustin}. Essentially, one can prove that the Fr\'{e}chet derivative of the mapping
$$
\sigma \mapsto \mbox{ values of } F(\sigma, u_i, \nabla u_i)
$$
(in appropriate function spaces) is an injective semi-Fredholm operator.

A variety of inversion procedures has been suggested and mostly tested on numerical phantoms: those involving numerical optimization \cite{Ammari_AET,Cap}, those reduced to solving well posed hyperbolic problems \cite{BalHybr}, or the ones that lead to solving transport equation or Poisson type elliptic equations \cite{KuKuAET,KuKuSynt}.

\begin{figure}[ht!]
       \includegraphics[scale=0.5]{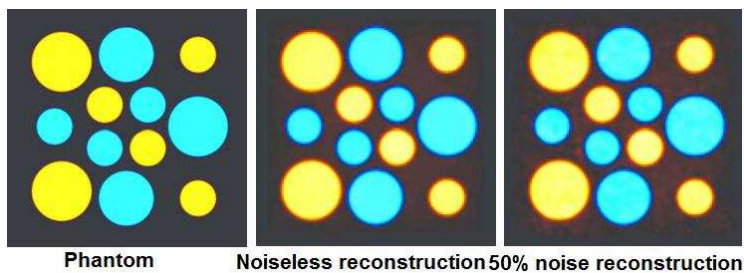}
\caption{An example of AET reconstruction. The picture is courtesy of L. Kunyansky.}\label{F:AET_rec}
\end{figure}

In most cases one could achieve wonderful quality reconstructions, e.g. see Fig. \ref{F:AET_rec}, where the method of \cite{KuKuAET} is used, which reduces to solving a Poisson equation for determining the conductivity.

Looking at Fig. \ref{F:AET_rec}, one might (and should) feel cheated. Indeed, how can one get such a good reconstruction with the data contaminated by a $50\%$ noise? The answer will be given in the next section.

By now, the mathematics of AET, albeit just a few years old, is already rather successful. The experimental implementation of the AET lags behind, due to the difficulty of acquiring good signal-to-noise ratios.

A different combination of ultrasound and EIT is suggested in \cite{ScherAET}. Here one again creates currents through the interior of the body of interest. These currents lead to a small inhomogeneous heating of the tissues, and thus to thermoelastic expansion. Then the TAT procedure, using the microphones surrounding the body, reconstructs a local functional $F(\sigma(x), u(x), \nabla u(x))$, after which one of the previously mentioned procedures of reconstruction can be applied.

\section{Synthetic focusing in hybrid techniques}\label{sec:focusing}

As it has been mentioned, the unfeasible \cite{Localized} perfect focusing is assumed in most mathematical work on AET, in particular, when using the small volume asymptotics. Can this be avoided? The answer, as it was explained in \cite{KuKuSynt} and then confirmed in \cite{KuKuAET}, is a ``yes.'' Indeed, the delta functions that are idealized focused beams, form a function ``basis.'' Suppose that we can produce another, practically feasible set of ultrasound waves, which would also form such a basis. Then, using the smallness of the acousto-electric effect, and thus linearity, one could mathematically process the data obtained from that basis of waves and ``synthetically focus'' them by changing the basis to delta-functions.

There are several examples of possible bases from \cite{KuKuSynt}:
\begin{itemize}
  \item Using large planar broad band transducers, one could generate a set of monochromatic planar waves with arbitrary wave vectors, and then the synthetic focusing would be just applying the inverse Fourier transform. This option was adopted in \cite{BalHybr}. Its practical feasibility is not yet clear.
  \item Using point-like omni-directional transducers, one could generate thin spherical shell waves. Then, lo and behold, the synthetic focusing will boil down to inversion of a restricted spherical mean transform, and thus any of the standard TAT inversions would do it. This is the option of \cite{KuKuAET}. The problem with this is that it is much easier to create a short $N$-shaped (Fig. \ref{F:N}) spherical wave rather than $\delta$-shaped such wave.
  \begin{figure}[ht!]
       \includegraphics[scale=0.8]{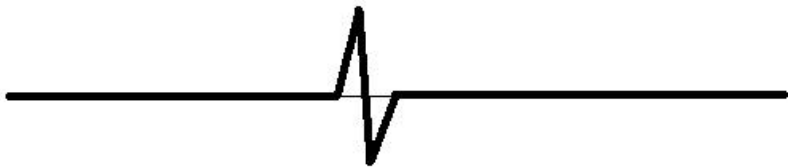}
\caption{An $N$-shaped pulse.}\label{F:N}
\end{figure}
  \item One can create narrow ``pencil beams'', as it is done in \cite{WangTextbook}, and then synthetic focusing would coincide with the inversion of the standard X-ray transform. This option has to struggle with impossibility of creating a truly homogeneous pencil ultrasound beam of sufficient length. \cite{Localized}
\end{itemize}

The sharp and amazingly stable with respect to the noise reconstruction shown in Fig. \ref{F:AET_rec}, was done using the second option for the basis: thin spherical waves, with consecutive TAT inversion for the synthetic focusing. After that an elliptic (Poisson type) equation was solved to recover the conductivity. Now, what about having $N$-shaped rather than $\delta$-shaped pulses? This ``difficulty'' turns out to be a blessing. Indeed, the TAT reconstruction includes a filtering portion, which increases the noise and decreases stability. However, with the $N$-shaped pulse, this filtration step is not needed, since it is already performed by the transducer. As the result, the synthetic focusing by the TAT inversion happens to be a \textbf{smoothing operator} and thus kills a lot of noise.
If we could indeed produce and use $\delta$-shaped pulses, the reconstruction would work, but would be very unlikely to survive a $50\%$ noise.

\section{Ultrasound Modulated Optical Tomography (UMOT)}\label{S:UOT}

The idea of scanning an object with a focused ultrasound that we applied in AET, can be tried with the optical tomography (OT) as well. The goal is the same: to improve drastically the resolution of OT (which is dismal on a centimeter depth and deeper). Since OT, like EIT, is a cheap, safe, and high contrast modality, achieving this goal, and thus adding high resolution, would make it an invaluable diagnostic tool.

In comparison with the AET, the situation with UMOT is reversed: there is an extensive body of experimental research (see \cite{WangTextbook} and references therein), but the first glimpses on the mathematics of UMOT are just appearing \cite{AllmBang,BalSchAOT,Nam,NamDob}, and even the mathematical model is not settled down.

The set-up of OT is as follows: One sends a beam of (coherent or incoherent) laser light through the body of interest and observes the intensity and speckle patterns of the outgoing light. The features of interest are the internal distribution of the absorption and scattering coefficients. The contrast in optical properties of cancerous and healthy locations is often huge. However, diffused photons, when they reach a detector after multiple scattering essentially loose any information about the locations they went through. This leads to dismal resolution at centimeters' depth (although good pictures can be obtained at skin depth). The idea of UMOT is to, concurrently with OT measurements, scan the body with a focused ultrasound and thus to acquire some interior (i.e., location-dependent) information, which hopefully would stabilize the problem. It has been argued in physics and biomedical engineering literature (e.g., \cite{WangTextbook,Kempe,Leutz}) that when using coherent light and measuring the ultrasound frequency Fourier component of the outgoing speckle pattern, one can recover the values of the following functional at an interior location $x$:
$$
G(x,d)A^2(x)I(x).
$$
Here $A$ is the applied ultrasound power (assumed to be known), $I$ is the light intensity, $d$ is the detector position on the boundary, and $G(x,d)$ is the ``probability of a photon emitted at the location $x$ to reach the detector at the location $d$''. In other words,  $G(x,d)$ is a Green's function of the diffusion equation
$$
-\nabla D(x) \nabla I(x) +\mu_a(x) I(x)=0
$$
inside the domain of the interest. The difficulty (at least for the author) is determining what the ``probability of reaching the detector'' means (e.g., does this mean the first time of reaching the point $d$?). Thus, it is not clear what boundary conditions the Green's function should correspond to.

It was assumed in \cite{AllmBang} that the correct boundary conditions are those that correspond to the optical impedance at the boundary of the object. Under this condition and with the perfect focusing assumption, a reconstruction algorithm was applied that showed sufficiently sharp internal reconstructions of the absorption coefficient $\mu_a$ (although the quality was lower than in the AET case). It was also shown in \cite{AllmBang} (see also the acknowledgments there) that (formally computed) Fr\'{e}chet derivative of the forward mapping is a semi-Fredholm operator in natural function spaces. However, injectivity of this derivative was not shown. Thus there are so far no local injectivity results.

Some controversy surrounds the usage of coherent light. It is claimed in engineering literature \cite{WangTextbook} that the signal from ultrasound modulation in the case of incoherent light could not be detected so far. However, there already are some mathematical studies of UMOT using incoherent light \cite{BalSchAOT}.

Synthetic focusing in UMOT is possible. However, while the spherical waves should still do the job, the use of planar waves and the consequent inversion of the Fourier transform seems to be not an option here, due to the presence of the \textbf{square} of the acoustic power $A(x)$ in the measured functional.

\section{Why the improvement? Inverse problems with interior information}\label{sec:interior}

The examples of AET and UMOT show how acquiring interior (i.e., attached to the internal locations) information stabilizes the utterly unstable inverse problems. This is an example of a folklore meta-statement: ``appropriate'' internal information stabilizes the severely unstable problems like diffused OT or EIT.

This issue was studied in detail in \cite{BalHybr}, where several internal information functionals arising in applications (including those described above) were studied. Different functionals required sometimes different techniques. There is a feeling though that there might be an answer to a
general question: What kind of a function $F(D(x), \sigma(x), u(x), \nabla u(x))$, if known, stabilizes the inverse boundary problems for
$$
-\nabla\cdot D(x) \nabla u+\sigma u=0?
$$

A rather general answer was just given in \cite{Dustin}. Under some reasonable conditions on the functional, which cover all cases we have considered here, it is shown that the Fr\'{e}chet derivative of the forward mapping is a semi-Fredholm operator. This explains why the observed improvement in stability occurs. Together with local uniqueness (which probably need to be proven individually in each case) this also gives local uniqueness and stability.

\section*{Acknowledgments}

This work was partially supported by the NSF DMS Grant 0604778, as well as by MSRI and IAMCS. The author expresses his gratitude to these institutions. Thanks also go to many colleagues with whom the author discussed the hybrid imaging methods (the author's attempt to write specific names produced a very long, while still incomplete list).


\begin{thebibliography}{99}
\bibitem{FR2} D. Finch  \& Rakesh,
Recovering a function from its spherical mean values in two and three dimensions, a Chapter in \cite{WangPATbook}, 77--88.
\end{thebibliography}
\end{document}